\documentstyle[12pt,twoside,fleqn,espcrc1,epsf]{article}

\newcommand{\AmS}{{\protect\the\textfont2
  A\kern-.1667em\lower.5ex\hbox{M}\kern-.125emS}}

\title{Higher twist light-cone distribution amplitudes of
vector mesons in QCD}

\author{Kazuhiro Tanaka\address{Department of Physics, Juntendo University, 
       Inba-gun, Chiba 270-1606, Japan}%
        \thanks{Supported in part by the Monbusho Grant-in-Aid for
Scientific Research No. 09740215.}
}

\begin{document}
\maketitle

\begin{abstract}
We present a systematic study of
twist-3 light-cone distribution
  amplitudes of vector mesons in QCD,
which is based on conformal expansion.
  A complete set of distribution amplitudes is constructed
  for $\rho$, $\omega$, $K^*$ and $\phi$ mesons,
which satisfies all (exact)
  equations of motion and constraints from conformal expansion.
\end{abstract}

\vspace{5mm}

Amplitudes of 
hard (light-cone dominated) exclusive processes in QCD
are expressed by the factorization formulae,
which separate the short-distance dynamics from the 
long-distance one\cite{BLreport}.
{}For 
those processes
producing a (light) vector meson $V$ ($= \rho, \omega, K^{*}, \phi$)
in the final state,
like exclusive semileptonic or radiative $B$ decays
($B \rightarrow Ve\nu$, $B\rightarrow V + \gamma$)
and the hard electroproduction of vector mesons 
($\gamma^{*} + N\rightarrow V + N'$),
a long-distance part involving the final vector meson
is given by the vacuum-to-meson matrix element of
the nonlocal
light-cone operators
$\langle 0| \bar{\psi}(z) \Gamma \lambda^{i} \psi(-z) |V \rangle$,
corresponding to universal nonperturbative quantity called 
the light-cone distribution amplitudes (DAs).
($z_{\mu}$ is a light-like vector
$z^{2} = 0$, and $\lambda^{i}$ and $\Gamma$ denote various
flavor and Dirac matrices.
We do not show the gauge phase factor connecting the quark 
and the antiquark fields.)

The DAs of higher twist are essential for systematic study
of preasymptotic corrections to hard exclusive amplitudes, and
are also interesting 
theoretically 
because they contain new and direct information
on hadron structure and the dynamics of QCD.

Here we will explicitly consider the ``chiral-odd'' DAs 
of the charged $\rho$ meson with momentum $P_{\mu}$
and polarization vector $e^{(\lambda)}_{\mu}$ 
($P^{2} = m_{\rho}^{2}$, $e^{(\lambda)}\cdot e^{(\lambda)} = -1$, 
$P\cdot e^{(\lambda)} = 0$).
The ``chiral-even'' DAs can be treated similarly\cite{BBKT98};
and extension to
other vector mesons is straightforward (see below).
The relevant quark-antiquark DAs are defined 
with the chirality-violating Dirac matrix structures 
$\Gamma = \{ \sigma_{\mu \nu}, 1\}$ as
\begin{eqnarray}
\langle 0| \bar{u}(z) \sigma_{\mu \nu} d(-z) 
| \rho^{-}(P, \lambda)\rangle
&=& if_{\rho}^{T} \left[
\left(e^{(\lambda)}_{\perp \mu}P_{\nu} -e^{(\lambda)}_{\perp \nu}P_{\mu} 
\right)
\int_{0}^{1} \!\! du\
e^{i\xi P\cdot z} 
\phi_{\perp}(u, \mu^{2})\right.
\nonumber \\
&+&\left. \left(P_{\mu}z_{\nu}-P_{\nu}z_{\mu}\right)
 \frac{e^{(\lambda)} \cdot z}{(P\cdot z)^{2}}m_{\rho}^{2}
\int_{0}^{1} \!\! du\
e^{i\xi P\cdot z} 
h_{\parallel}^{(t)}(u, \mu^{2}) 
\right],
\label{eq:defT}
\end{eqnarray}
\begin{equation}
\langle 0| \bar{u}(z) d(-z) 
| \rho^{-}(P, \lambda)\rangle
= -i
\left(
f_{\rho}^{T}
- f_{\rho}\frac{m_{u}+m_{d}}{m_{\rho}}\right)
m_{\rho}^{2}
(e^{(\lambda)} \cdot z)  
\int_{0}^{1} \!\! du\
e^{i\xi P\cdot z}
h_{\parallel}^{(s)}(u, \mu^{2}).
\label{eq:defS}
\end{equation}
In (\ref{eq:defT}),
we neglect the Lorentz structures corresponding to 
the twist-4 terms 
for simplicity (see Ref.\cite{BBKT98} for the complete expressions).
Our Lorentz frame is chosen as $P\cdot z = P^{+}z^{-}$.
The nonlocal operators on the l.h.s.
is renormalized at scale $\mu$.
We set $\xi \equiv u - (1-u) = 2u-1$,
and $f_{\rho}$ and $f_{\rho}^{T}$ are 
the usual vector and tensor decay constants as
$\langle 0| \bar{u}(0) \gamma_{\mu} d(0) 
| \rho^{-}(P, \lambda)\rangle = f_{\rho} m_{\rho} e^{(\lambda)}_{\mu}$
and
$\langle 0| \bar{u}(0) \sigma_{\mu \nu} d(0) 
| \rho^{-}(P, \lambda)\rangle = i f_{\rho}^{T}(e^{(\lambda)}_{\mu}P_{\nu}
- e^{(\lambda)}_{\nu}P_{\mu})$.
We wrote 
$e^{(\lambda)}_{\mu} = e^{(\lambda)}_{\parallel \mu} 
+ e^{(\lambda)}_{\perp \mu}$
with $e^{(\lambda)}_{\parallel \mu} = [P_{\mu} 
- z_{\mu}m_{\rho}^{2}/(P\cdot z)](e^{(\lambda)}\cdot z)/(P\cdot z)$,
so that the DAs with the subscripts $\parallel$ and $\perp$
describe longitudinally
and transversely 
polarized $\rho$ mesons, respectively.
$\phi_{\perp}$ is of twist-2, while 
$h_{\parallel}^{(t)}$ and $h_{\parallel}^{(s)}$ are of twist-3.
These DAs are dimensionless functions
and describe the probability amplitudes
to find the $\rho$ in a state with quark and antiquark,
which carry the light-cone momentum fractions $u$ and
$1-u$, respectively,
and have a small transverse separation of order $1/\mu$.


The 
quark-antiquark-gluon DAs
can be defined
similarly;
they are of twist-3 and higher. There exists one chiral-odd
DA of twist-3, which is given by
\begin{eqnarray}
\lefteqn{
\langle 0| \bar{u}(z) \sigma^{\mu \nu}z_{\nu} gG_{\mu\eta}(vz)
z^{\eta}
d(-z) | \rho^{-}(P, \lambda)\rangle}\nonumber\\
&&\;\;\;\;\;\;\;\;\;\;= 
(P\cdot z)(e^{(\lambda)} \cdot z)f_{3\rho}^{T}m_{\rho}
\int_{0}^{1} 
\!\!{\cal D}\underline{\alpha}\
e^{-iP\cdot z(\alpha_{u} - \alpha_{d} + v\alpha_{g})} 
{\cal T}(\underline{\alpha}, \mu^{2}) ,
\label{eq:def3DA}
\end{eqnarray}
where $G_{\mu \eta}$ is the gluon field strength tensor,
${\cal D}\underline{\alpha} \equiv d\alpha_{d}d\alpha_{u} d\alpha_{g}
\delta(1 - \alpha_{d}-\alpha_{u}-\alpha_{g})$, and $\underline{\alpha}$
denotes the set of three light-cone momentum fractions: 
$\alpha_{d}$ (quark), $\alpha_{u}$ (antiquark), and $\alpha_{g}$
(gluon).
$f_{3\rho}^{T}$ is the three-body tensor decay constant\cite{BBKT98},
so that ${\cal T}$ is dimensionless
and is conveniently normalized as
$\int {\cal D}\underline{\alpha} (\alpha_{d} - \alpha_{u}) 
{\cal T}(\underline{\alpha}) = 1$.
The three-particle DA ${\cal T}$ describes a higher component
in the Fock-state wave function with additional gluon.

The basis of DAs defined above is overcomplete
due to the constraints from the QCD equations of motion.
Using 
$(i 
\not{\!\! D} - m_{q})q(x) =0$ ($q= u, d$), 
we obtain 
\begin{eqnarray}
\lefteqn{\frac{1}{2} x^{\nu} \frac{\partial}{\partial x_{\mu}}
\bar{u}(x) \left[ \gamma_{\mu}, \gamma_{\nu}\right]_{\pm}
d(-x)
= \int_{-1}^{1}\! dv v^{(1 \mp 1)/2}
\bar{u}(x) \sigma^{\mu \nu} x_{\nu}
g G_{\mu \eta}(vx) x^{\eta} d(-x)} \nonumber \\
&&\;\;\;\;\;\;\;\;\;\;\;\;\;\;\;\;\;\;\;\;
- \frac{x^{\nu}}{2}\frac{\partial}{\partial y_{\mu}}
\left. \left\{ \bar{u}(x + y) 
\left[ \gamma_{\mu}, \gamma_{\nu}\right]_{\mp}
d(-x + y) \right\}\right|_{y \rightarrow 0}
\!\! + i (m_{u} \pm m_{d}) \bar{u}(x) \rlap/{\mkern-1mu x} d(-x),
\label{eq:id}
\end{eqnarray}
where 
$\left[ \gamma_{\mu}, \gamma_{\nu} \right]_{\pm}
\equiv \gamma_{\mu}\gamma_{\nu} \pm \gamma_{\nu}\gamma_{\mu}$.
In the light-cone limit $x^{2} \rightarrow 0$,
the vacuum-to-meson matrix elements of (\ref{eq:id})
yield a system of integral equations
between two- and three-particle DAs defined above.
We note that the 
total derivative term
induces mixing between $h_{\parallel}^{(t)}$ 
and $h_{\parallel}^{(s)}$.
We can solve\cite{BBKT98} these coupled integral equations 
in a form ($i=s, t$)
\begin{equation}
h_{\parallel}^{(i)}(u, \mu^{2}) =
\int_{0}^{1} \!dv\ K^{(i)}_{WW}(u,v) \phi_{\perp}(v, \mu^{2}) 
+ \int_{0}^{1}\! {\cal D}\underline{\alpha}\ 
K^{(i)}_{g}(u, \underline{\alpha})
{\cal T}(\underline{\alpha}, \mu^{2}),
\label{eq:sol}
\end{equation}
where $K^{(i)}_{X}$ ($X=WW, g$) are independent of $\mu$. 
We omit
the quark mass correction term for simplicity.
The first term on the r.h.s. is 
the twist-2 contribution and 
thus the analogue of the 
Wandzura-Wilczek piece of the nucleon structure function 
$g_{2}(x, Q^{2})$. 

To analyze (\ref{eq:sol}),
we introduce the conformal partial wave expansion
in the $m_{q}\rightarrow 0$ limit\cite{BBKT98,BF90}.
The conformal expansion of light-cone DAs
is analogous to the partial wave expansion
of wave functions in standard quantum mechanics (QM).
In conformal expansion,
the invariance of massless QCD under conformal transformations
substitutes the rotational symmetry in QM.
In QM, the purpose of partial wave
decomposition is to separate angular degrees of freedom
from radial ones
(for spherically symmetric potentials). 
All dependence on the angular
coordinates is included 
in spherical harmonics which form an irreducible
representation of the group O(3), 
and the dependence on the single
remaining radial coordinate is governed by a one-dimensional
Schr\"{o}dinger equation. 
Similarly, the conformal expansion of DAs in QCD aims 
to separate longitudinal degrees of freedom
from  transverse ones. 
All dependence on the longitudinal momentum fractions
is included in terms of certain orthogonal polynomials
which form irreducible representations 
of the so-called collinear subgroup
of the conformal group, SL(2,R). 
The transverse-momentum dependence
(the scale-dependence) is governed 
by simple renormalization group equations:
the different partial waves,
labeled by different ``conformal spins'',
behave independently and do not mix with each other.
Since the conformal invariance of QCD is
broken by quantum corrections, mixing of different terms of the
conformal expansion is only absent to leading logarithmic accuracy.
Still, conformal spin is a good quantum number in hard processes,
up to small corrections of order $\alpha_{s}^{2}$.

The conformal expansion of the DAs on the r.h.s. of
(\ref{eq:sol}) reads
\begin{equation}
\phi_{\perp}(u, \mu^{2}) = 6u\bar{u} \sum_{n=0}^{\infty} 
a_{n}(\mu^{2}) C_{n}^{3/2}(\xi); \;\;\;\;\;\; 
{\cal T}(\underline{\alpha}, \mu^{2}) 
= 360 \alpha_{d}\alpha_{u}\alpha_{g}^{2} \sum_{k,l=0}^{\infty}
\omega_{kl}(\mu^{2}) J_{kl}(\alpha_{d}, \alpha_{u}),
\label{eq:ce2}
\end{equation}
where $\bar{u}= 1-u$, $\xi=2u-1$, 
$\alpha_{g} = 1 - \alpha_{d} - \alpha_{u}$, and $C_{n}^{3/2}$ and
$J_{kl}$ are particular Gegenbauer and Appell 
polynomials\cite{BBKT98}.
Using orthogonality relations of these polynomials,
the expansion coefficients are expressed by matrix elements 
of local conformal operators:
$a_{n}$ and $\omega_{kl}$ are given by 
local operators of conformal spin $j= n+2$ and $j= k+l+7/2$,
respectively
(three-particle conformal 
representations are degenerate).
Thanks to conformal symmetry,
$K_{X}^{(i)}$ of (\ref{eq:sol}) are
resolved into the superposition of the terms of definite
conformal spin.
As a result, $h_{\parallel}^{(i)}$ are given by the conformal
expansion, where each expansion coefficient 
is expressed in terms of $a_{n}$ and $\omega_{kl}$ 
with the corresponding spin $j$. 

Now 
all DAs are expressed,
order by order in the conformal expansion,
by independent matrix elements
$a_{n}$ and $\omega_{kl}$.
{}From (\ref{eq:sol}) and (\ref{eq:ce2}),
the $\mu^{2}$-dependence of the DAs is governed by that
of $a_{n}(\mu^{2})$ and $\omega_{kl}(\mu^{2})$.
This is determined by the renormalization of the corresponding
local conformal operators, and
is worked out in the leading logarithmic 
approximation\cite{BBKT98,KNT98}.
The results indicate that the relevant anomalous dimensions
increase as $\sim \ln j$. 
This means that only the first few conformal partial waves 
contribute at sufficiently large scales.
Therefore, the truncation of the conformal expansion 
at some low order provides a useful and consistent 
approximation of the full DAs.

\begin{figure}[t]
\leftline{\epsffile{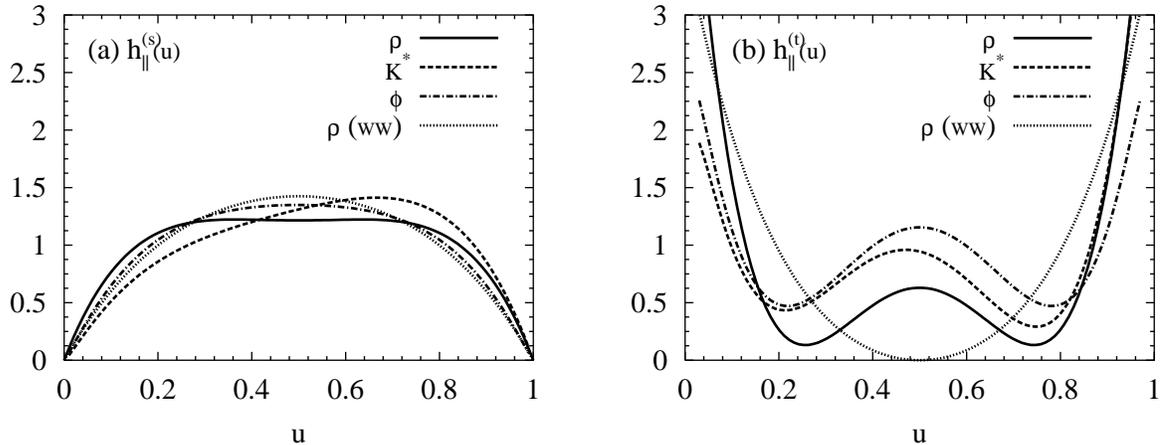}}
\vspace{-3.4cm}
\caption[]{Two-particle twist-3 DAs for the
$\rho$, $K^*$ and $\phi$ mesons:
(a) $h_{\parallel}^{(s)}(u)$ of (\ref{eq:trun});
(b) $h_{\parallel}^{(t)}(u)$ of (\ref{eq:trun2}).
``$\rho$ (WW)'' denotes the ``Wandzura-Wilczek'' type
contribution given by the first three 
terms of (\ref{eq:trun}) and (\ref{eq:trun2}) 
for the case of the $\rho$ meson.}\label{fig:1}
\end{figure}

We take into account the partial waves with $j \le 9/2$,
where the terms with $n \le 2$ and $k+l \le 1$ 
are retained in 
the first and second equations of (\ref{eq:ce2}).
Correspondingly, we get\cite{BBKT98}
\begin{eqnarray}
h_{\parallel}^{(s)}(u) & = & 
6u\bar{u} \left[ 1 + a_1 \xi + \frac{1}{4}a_2 (5\xi^2-1)\right]
+ 35 u\bar{u}\zeta_{3} (5\xi^2-1) \nonumber\\
&+& 3 \delta_{+} (3 u \bar{u} + \bar{u} \ln \bar{u} + u \ln u) 
+ 3 \delta_{-}  (\bar{u}
\ln \bar{u} - u \ln u), \label{eq:trun}\\
h_{\parallel}^{(t)}(u) &= & 
3\xi^2+ \frac{3}{2}a_1 \xi (3 \xi^2-1)
+ \frac{3}{2} a_2 \xi^2 (5\xi^2-3)
+\frac{35}{4}\zeta_{3}(3-30\xi^2+35\xi^4) \nonumber\\
&+& \frac{3}{2} \delta_+
(1 + \xi \ln \bar{u}/u) + \frac{3}{2}\delta_- \xi ( 2
+ \ln u + \ln\bar{u} ).
\label{eq:trun2}
\end{eqnarray}
Here $\zeta_{3} = f_{3\rho}^{T}/(f_{\rho}^{T}m_{\rho})$,
and we incorporated
the quark mass corrections proportional to 
$\delta_{\pm} = f_{\rho}(m_{u}\pm m_{d})/(f_{\rho}^{T} m_{\rho})$.
The results for other vector mesons $\omega$, $K^{*}$ and $\phi$
are obtained by trivial substitutions.
We emphasize that these results provide a consistent set
of DAs involving a minimum number of 
independent nonperturbative parameters.
These parameters are calculated from QCD sum rules
taking into account SU(3)-breaking effects\cite{BBKT98}.
The resulting DAs are plotted in Fig.~\ref{fig:1}.
The comparison between the solid and dotted curves 
shows that the three-particle contributions
(the term with $\zeta_{3}$ in (\ref{eq:trun}), (\ref{eq:trun2}))
are important and broaden the distributions.
On the other hand, 
the quark mass effects are not so large except for the end-points
$u \rightarrow 0$ and $u\rightarrow 1$ for $h_{\parallel}^{(t)}(u)$.

In conclusion, we have developed a powerful framework
for hard exclusive processes,
which allows 
to express the higher twist DAs
in a model-independent way by a minimum number of nonperturbative
parameters. Combined with estimates of the nonperturbative parameters,
this leads to  
model building of the DAs consistent with all 
QCD constraints. Our formalism is applicable
to arbitrary twist\cite{BB982} and other light mesons\cite{B98}, 
and our results are immediately
applicable to a range of phenomenologically 
interesting processes\cite{BB98}.

%
%

\bigskip
The author would like to thank P. Ball, V.M. Braun, and
Y. Koike for the collaboration on the subject discussed 
in this work.

\end{document}